\documentclass[%
 preprint,
 amsmath,amssymb,
 aip,
]{revtex4-2}

\usepackage{booktabs}
\usepackage{diagbox}

\usepackage{graphicx} 
\usepackage{appendix}
\usepackage{physics}

\usepackage{multirow}
\usepackage{xcolor}

\usepackage{amsmath}

\usepackage{xr}
\usepackage{cleveref}

\usepackage{tikz}
\usetikzlibrary{intersections,backgrounds}

\graphicspath{{../FitData}}
\begin{document}

\newcommand{\eqn}[1]{Eq.\ (\ref{#1})}
\newcommand{\eqs}[2]{Eqs.\ (\ref{#1}--\ref{#2})}
\def\br{{\mathbf{r}}}
\def\brp{{\mathbf{r}^\prime}}
\def\bz{{\mathbf{z}}}
\def\bzp{{\mathbf{z}^\prime}}

\renewcommand{\mp}[1]{\textcolor{red}{#1}}

\author{Valeria Rios-Vargas}
\email{valeria.rios@rutgers.edu}
\affiliation{Department of Physics, Rutgers University, Newark, NJ 07102, USA}

\author{Ezekiel Oyeniyi}
\affiliation{Department of Physics, University of Ibadan, Ibadan, Nigeria}

\author{Xuecheng Shao}
\affiliation{Key Laboratory of Material Simulation Methods \& Software of Ministry of Education, College of Physics, Jilin University, Changchun 130012, PR China}

\author{Wala Fathelrahman Ibrahim Elsayed}
\affiliation{Condensed Matter Section, East African Institute for Fundamental Research (EAIFR), University of Rwanda,  Kigali, Rwanda.}

\author{Sunday Joseph Ogenyi}
\affiliation{Condensed Matter Section, East African Institute for Fundamental Research (EAIFR), University of Rwanda,  Kigali, Rwanda.}

\author{Alex Okello}
\affiliation{Department of Science and Vocational Education, Faculty of Education, Lira University, Lira, Uganda}

\author{Michele Pavanello}
\email{m.pavanello@rutgers.edu}
\affiliation{Department of Physics, Rutgers University, Newark, NJ 07102, USA}
\affiliation{Department of Chemistry, Rutgers University, Newark, NJ 07102, USA}

\date{\today}

\begin{abstract}
Developing reliable pseudopotentials for orbital-free density functional theory (OF-DFT), especially for transition metals, remains a significant challenge. In this study, we provide a theoretical framework for analyzing pseudization strategies for OF-DFT calculations. From the analysis arises a proposed pseudization method which involves constructing local pseudopotentials by targeting existing Kohn–Sham DFT pseudopotentials through an optimized effective potential procedure. We produce four distinct sets of local pseudopotentials and evaluate their accuracy and transferability on the transition metal elements. Our results indicate a substantial improvement over previously available pseudopotentials. Although current OF-DFT functionals still only reach a qualitative accuracy for transition metals, our newly developed pseudopotentials provide a rigorous framework for further methodological advancements.
\end{abstract}

\title{Pseudopotentials for Orbital-Free DFT: Capturing Nonlocality and Correcting Functional Approximants}
\maketitle

\section{Introduction and Background}
Among all approaches to modeling materials at the atomic level, Density Functional Theory as formulated by Kohn and Sham \cite{kohn1965self} (KS-DFT) stands out for its impact. Its most widely used methods have so far been cited over 200,000 times\cite{perdew1996generalized}.  KS-DFT's implementations offer a good compromise between accuracy and computational cost. However, independently from the choice of the underlying approximations, it becomes too computationally expensive when large, realistically-sized systems (e.g., quantum dots, proteins and complex materials interfaces) are targeted. The reason for this is the unavoidable costs of building and diagonalizing the KS Hamiltonian matrix. In particular, diagonalizations carry a cost of $\mathcal{O}(N^3)$ number of operations, where $N$ is measure of system size. There are many approaches that tackle KS-DFT's steep computational scaling \cite{liou2020scalable,dawson2022density,miyazaki2022large,xue2024accelerating}. Among them, a promising alternative is orbital-free DFT (OF-DFT) \cite{mi2023orbital,witt2018orbital}. 

OF-DFT is based on the same theoretical footing as KS-DFT \cite{kohn1965self}. In particular, it borrows KS-DFT's expression for the total electronic energy given as a functional of the electronic density:
\begin{equation}
    E[n] = T_s [n] + E_{Hxc}[n] +  \int v_{ext}(\br) n(\br) d\br
    \label{eq:toten}
\end{equation}
where $T_s$, $E_{Hxc}$, and $v_{ext}(\br)$ are the noninteracting kinetic energy, the Hartree exchange-correlation (xc) energy, and the interaction with the external potential (typically the Coulomb electron-nuclear attraction potential), respectively. The noninteracting kinetic energy is given in terms of KS orbitals (single particle functions, hence the ``s'' subscript) as follows
\begin{equation}
    \label{eq:ts}
    T_s[n]\equiv T_s[\{\phi_i[n]\}] = -\frac{1}{2}\sum_i n_i \langle \phi_i | \nabla^2 | \phi_i \rangle,
\end{equation}
where $n_i$ are occupation numbers. The electron density is simply
\begin{equation}
    \label{eq:dens}
    n(\br) = \sum_i n_i |\phi_i(\br)|^2
\end{equation}
The Hartree functional is defined as the electronic Coulomb self-repulsion, $E_H[n]=\frac{1}{2}\int \frac{n(\br)n(\brp)}{|\br-\brp|}d\br d\brp$ and the xc functional, $E_{xc}[n]$ is the energy term that is added to $T_s[n]$, $E_H[n]$ and $\int v_{ext}(\br)n(\br)d\br$ to recover the total electronic energy.

Following from \eqn{eq:dens}, we can use the KS orbitals as variational functions to minimize the total energy, leading to the canonical KS equations
\begin{equation}
    \label{eq:kseq}
    -\frac{1}{2}\nabla^2 \phi_i(\br) + v_s(\br)\phi_i(\br) = \varepsilon_i \phi_i(\br)
\end{equation}
where $v_s(\br) =  v_{Hxc}[n](\br) + v_{ext}(\br)$ defining the Hxc potential as a functional derivative, $v_{Hxc}[n](\br) = \frac{\delta E_{Hxc}}{\delta n(\br)}$.

In OF-DFT, the noninteracting kinetic energy, $T_s$, is evaluated approximately by a pure density functional, which we call henceforth the kinetic energy density functional or KEDF. 
 Using KEDFs allows the energy minimization of the functional in \eqn{eq:toten} to be based solely on the electron density, completely bypassing the KS orbitals, and generally leading to algorithms of linear scaling complexity \cite{mi2023orbital}. The low computational scaling resulting from the employment of approximate KEDFs comes at an accuracy cost. In fact OF-DFT development mostly focuses on development of KEDFs which are typically either semilocal \cite{laricchia2011generalized,constantin2011semiclassical,luo2018simple} or nonlocal \cite{huang2010nonlocal,shao2021revised,mi2019orbital}. In this work, we will not focus on KEDF development but we recognize that the accuracy of any result in this work is heavily biased by the quality of the KEDF employed.

One thorn on the side of OF-DFT has persisted. Namely the handling of the electron-ion interaction. While this seems a non-issue (the $v_{ext}(\br)$ is conceptually trivial), 
when approaching condensed phases the pseudopotential (PP) approximation is usually invoked \cite{wang2003first, watson1998ab, legrain2015highly, chai2007orbital, chi2024high, mi2016first,ke2013angular}. Natural basis functions used to expand the KS orbitals for materials are plane waves, $\chi_{\mathbf{k}}(\br) = \frac{1}{\sqrt{\Omega}}e^{i\mathbf{k}\cdot \br}$, where $\mathbf{k}$ denotes a wave vector in the reciprocal lattice, and $\Omega$ is the real-space volume of the primitive cell such that the plane waves are orthonormal over $\Omega$. To maintain a reasonable number of plane waves, a cutoff is imposed $|\mathbf{k}|<k_{max}$. Thus, PPs achieve two objectives: (1) at the expense of introducing nonlocality in the pseudopotential, they remove the need to explicitly consider core electrons as these are, in most applications, chemically inert; and (2) they eliminate the singularity of the Coulomb electron-nuclear potential at the nuclear positions \cite{garrity2014pseudopotentials, hamann1979norm}.

Unfortunately, accurate and transferable nonlocal PPs (NLPPs) commonly used in KS-DFT, are not applicable to OF-DFT because they require KS orbitals to couple to the nonlocal part of the PP. In OF-DFT, as we have seen, orbitals are not immediately available.

Although we refer to recent reviews for additional details \cite{mi2023orbital,witt2018orbital}, in the past decades several attempts have produced effective local PPs (LPPs, hereafter) that have opened OF-DFT's applicability to main group metal elements, groups III-V elements third row and below, and a selected number of transition metal elements (such as Ag) \cite{mi2016first,huang2008transferable,zhou2004transferable,del2017globally,legrain2015highly}. 

The guiding principles to build LPPs have been to employ them in a KS-DFT calculation (i.e., with exact $T_s$) imposing it to reproduce the electron density, lattice constant and even the band structure of solids \cite{mi2016first,huang2008transferable,zhou2004transferable}. For those elements for which this is possible, it is said LPPs can be employed in OF-DFT calculations.  It is certainly surprising that band structures can be reproduced in pseudopotential calculations having only a local part. However, in many cases it was shown to be possible \cite{starkloff1977local}. It turns out that LPPs can be successfully constructed for the main-group elements mentioned above (provided that the semicore states are excluded from the pseudization \cite{chi2024high}) and also for a large array of transition metals (where instead semicore electrons are included). The most accurate LPPs available to date for transition metal elements are the recently developed highly accurate LPPs by the Huang group \cite{chi2024high}. The valence density of transition metals consists of contributions from both localized $d$ states and delocalized $s$–$p$ states. Because existing KEDFs are based on the response of the homogeneous or nearly homogeneous electron gas, they struggle to provide accurate kinetic energies for transition metals \cite{huang2012toward}. In addition, accurate pseudizations of transition metals should account for semicore states that lie close in energy to the valence $d$ bands and influence bonding and screening. Since OF-DFT represents $T_s[n]$ as a pure functional of the electron density, it lacks explicit angular-momentum dependence \cite{ke2013angular} and therefore cannot distinguish $d$-state character from $s$ or $p$ contributions. These combined factors explain why transition metals are a particularly demanding class of atoms for OF-DFT and motivate the construction of pseudopotentials that mitigate these limitations.

A flaw, however, is that the LPPs have so far been developed based on comparison between KS-DFT calculations with LPPs and KS-DFT calculations with NLPPs. When the band structure and other observables agree between the two, the LPP is deemed suitable for OF-DFT simulations. However, this procedure does not account for the differences between OF-DFT KEDFs and the exact $T_s[n]$ from KS-DFT. In this work we aim at addressing this flaw.

In fact, early on it was recognized that LPPs should be constructed specifically for approximate OF-DFT functionals so that they could not only provide a much needed smooth electron-ion interaction potential, but they could also provide partially a correction to the approximate KEDF potential \cite{watson1998ab}. These LPPs were soon abandoned \cite{zhou2004transferable} because the resulting LPPs had features that would induce numerical instability, would not lead to proper Coulomb decay of the LPP and, most importantly, were found to yield mixed results for condensed phases. So it seems we are at an impass, on one hand employing KS-DFT to build LPPs is likely to generate LPPs that are of little use in practical OF-DFT simulations where approximate KEDFs are used. On the other hand, building LPPs based on approximate KEDFs is associated with unmanageable side effects.

We also mention a notable recent approach, which involves evaluating directly the nonlocal part of the NLPP with orbital-free one-body reduced density matrices \cite{xu2022nonlocal}. While the idea is promising, the resulting pseudopotentials are only accurate for a limited set of atoms, pointing to the fact that orbital-free formulations of reduced density matrices are often not $N$-representable \cite{chakraborty2017two}.

Considering the pros and cons of the various LPP design principles, in this work we aim to produce a library of LPPs for transition metal elements that can be used in conjunction with approximate KEDFs. The design principles adopted in this work arise from a careful formal analysis of atomic pseudization (carried out in the following section) and can be summarized with three points: (1) LPPs should target an \textit{existing} set of NLPPs, (2) LPPs should differ from the NLPP's local part by a correction that is short ranged, and finally (3) LPPs should be constructed in conjunction with approximate KEDFs. 

\section{An in-depth look at pseudopotentials and pseudodensities}

In the PP approximation, the core electrons of each atom are represented by an effective nonlocal potential and the KS equations are only invoked for the valence electrons. The external potential becomes
\begin{align}
    \label{eq:kspot}
   v_{ext}(\br) \to v_{NLPP}(\br,\brp) &= v_{loc}(\br) \delta(\br-\brp)+ v_{nl}(\br,\brp) \\
   \nonumber 
   &\equiv v_{loc}(\br) + \hat v_{nl}
\end{align}
where we have indicated with a hat the nonlocal part of the NLPP.

NLPPs give rise to an electronic density $n(\mathbf{r})$ (called pseudodensity) that only reflects the density of the valence electrons outside the core radii of the atoms and is smooth inside. Such radii (NLPP cutoff radii, or $R_{cut}^{nl}$) are parameters of the NLPPs and their value is determined by convenience. Small $R_{cut}^{nl}$ will yield more accurate NLPPs but at the expense of needing to represent the more oscillatory components of the valence electrons.

The physical meaning of the nonlocal part of the NLPP is to maintain orthogonality between the valence and core orbitals. In KS-DFT, $\hat v_{nl}$ is usually given in terms of a limited number of spherical harmonic projectors centered on the atoms. 

\subsection{Nonlocal Pseudopotential Representations}

Once a particular NLPP is introduced, the solution of the KS equations yields a pseudodensity, $n(\br)$. That same pseudodensity can be represented by many other NLPPs (hereafter indicated by $\hat v_{NLPP}$). The usual map connecting external {\it potentials} with {\it densities} applies. However, this time we connect external, possibly nonlocal  {\it pseupotentials} with a target {\it pseudodensity} 
\begin{equation}
    \label{map:nlpp}
    \hat v_{NLPP} \longrightarrow n(\br).
\end{equation}

Hohenberg and Kohn theorems for nonlocal external potentials due to Gilbert \cite{gilbert1975hohenberg} apply and guarantee the existence of the above map, even though it is generally not bijective. The non-bijectivity is due to the fact that for a given choice of $\hat v_{nl}$ one can always find an associated $v_{loc}(\br)$ such that the KS equations yield the target density $n(\br)$. Therefore there are multiple (potentially an infinite number) of $\hat v_{NLPP}$ that have nonzero nonlocal part and that yield the same electron density, $n(\br)$. We call the set of such NLPPs a {\it NLPP representation}. Typically, there is one NLPP having the property that beyond $R_{cut}^{nl}$ not only the pseudodensity matches the all-electron density, but also the pseudo KS wavefunctions match the all-electron KS wavefunction for the isolated atom \cite{kleinman1982efficacious}.

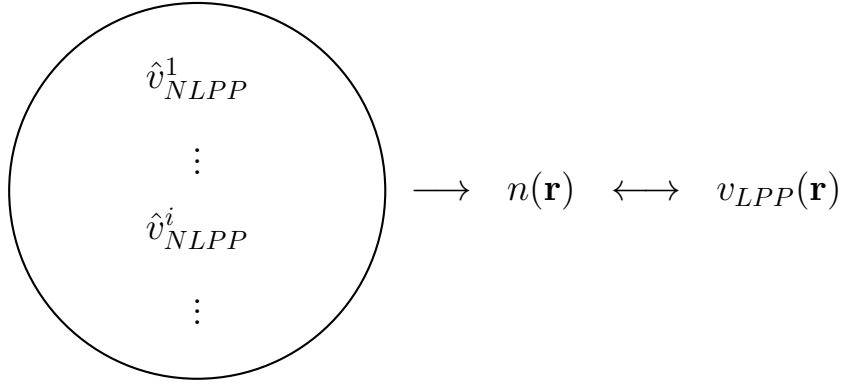
\begin{figure}
    \centering
    \begin{tikzpicture}
        \begin{scope}
            \clip (0,0) circle(2.5);
            \fill[white] (1.4,0) circle(2.5);
        \end{scope}
        \draw[thick] (0,0) circle(2.5);
        \node at (0,1.5) {\large $\hat v_{NLPP}^1$};
        \node at (0,0.5) {\large $\vdots$};
        \node at (0,-0.5) {\large $\hat v_{NLPP}^i$};
        \node at (0,-1.5) {\large $\vdots$};
        \node at (5.7,0) {\large $\longrightarrow ~~ n(\br) ~~ \longleftrightarrow ~~ v_{LPP}(\br)$};
    \end{tikzpicture}
\caption{Depiction of the maps in Eqs.\ (\ref{map:nlpp},\ref{map:lpp}) and the concept of NLPP representation (set of nonlocal PPs on the left-hand side) mapping to a target pseudodensity, $n(\br)$. The bijectivity of the map connecting the pseudodensity with the associated local KS potential, $v_{LPP}(\br)$ is indicated by a left-right arrow. } \label{fig:maps}
\end{figure}

The Hohenberg and Kohn theorems guarantee that if $n(\br)$ is noninteracting $v_s$-representable, then there is one and only one purely local PP (hereafter denoted by $v_{LPP}(\br)$) that, employed as external potential in the KS equations, yields the pseudodensity, $n(\br)$. Namely, the bijective map
\begin{equation}
    \label{map:lpp}
    v_{LPP}(\br) \longleftrightarrow n(\br)
\end{equation}
exists. We depict the concept of NLPP representation and the maps in Eqs.\ (\ref{map:nlpp}, \ref{map:lpp}) in Figure \ref{fig:maps}.

The relation between the local PP, $v_{LPP}(\br)$, and any of the NLPPs, $\hat v_{NLPP}^i$, can be derived on the basis of the optimized effective potential method (OEP) \cite{gorling2005orbital,yang2002direct}, as we now explain.

\subsection{Relating Local and Nonlocal Pseudopotentials}

We now seek to find the local PP, $v_{LPP}(\br)$, that yields the target pseudodensity, $n(\br)$. We first define the relevant energy functionals, one for a NLPP and one for a LPP,

\begin{align}
    \label{eq:NLPPKS}
    E_{\hat v_{nl}}[n] &= T_s^{\hat v_{nl}}[n] + E_{NL}^{\hat v_{nl}}[n] + E_{Hxc}[n] + \int v_{loc}^{\hat v_{nl}}(\br)n(\br) d\br\\
    \label{eq:LPPKS}
    E[n] &= T_s[n] + E_{Hxc}[n] + \int v_{LPP}(\br)n(\br) d\br
\end{align}
where we have indicated with a superscript the quantities that are dependent on the choice of the purely nonlocal part of the NLPP, and we define the noninteracting kinetic energy in the NLPP case as 
\begin{equation}
    \label{eq:tsnl}
    T_s^{\hat v_{nl}}[n] = -\frac{1}{2}\sum_i n^{\hat v_{nl}}_i \langle \phi^{\hat v_{nl}}_i | \nabla^2| \phi^{\hat v_{nl}}_i \rangle
\end{equation}
and the nonlocal energy, $E_{NL}^{\hat v_{nl}}[n]$,
\begin{equation}
    \label{eq:enl}
    E_{NL}^{\hat v_{nl}} = \sum_i n^{\hat v_{nl}}_i \langle \phi^{\hat v_{nl}}_i | \hat v_{nl}| \phi^{\hat v_{nl}}_i \rangle = {\rm Tr}\left[ \hat v_{nl} \hat \gamma_s^{\hat{v}_{nl}}\right].
\end{equation}
where $\gamma_s^{\hat{v}_{nl}}$ is the KS density matrix of the NLPP KS system.

Crucially, we must require that $E_{\hat v_{nl}}[n] = E[n]$, Which delivers an important relation,
\begin{equation}
    \label{eq:NL2L}
    T_s^{\hat v_{nl}}[n] + E_{NL}^{\hat v_{nl}}[n] + \int v_{loc}^{\hat v_{nl}}(\br)n(\br) d\br = T_s[n] + \int v_{LPP}(\br)n(\br) d\br
\end{equation}
which can be rearranged to
\begin{equation}
    \label{eq:NL2L}
    T_s^{\hat v_{nl}}[n] + E_{NL}^{\hat v_{nl}}[n]  = T_s[n] + \int \Delta v_{LPP}(\br)n(\br) d\br.
\end{equation}
In the above, we defined $\Delta v_{LPP}(\br) = v_{LPP}(\br) - v_{loc}^{\hat v_{nl}}(\br)$, as the potential that achieves a KS inversion \cite{jacob2011unambiguous,jensen2018numerical,yang2002direct} of the NLPP KS system onto a LPP KS system.

Taking a functional derivative yields a crucial result
\begin{equation}
    \label{eq:inv}
    \Delta v_{LPP}(\br) =\frac{\delta T_s^{\hat v_{nl}}}{\delta n(\br)} - \frac{\delta T_s}{\delta n(\br)} + \frac{\delta E_{NL}^{\hat v_{nl}}}{\delta n(\br)}
\end{equation}

The above functional derivatives involve orbital-dependent functionals and thus can only be approached with OEP methods \cite{gorling2005orbital}. In the supplementary information document, we show how to apply OEP in this setting and we also show that $\frac{\delta E_{NL}^{\hat v_{nl}}}{\delta n(\br)}$ decays faster than the Coulomb interaction, specifically like a dipole field outside $R_{cut}^{nl}$. 

In addition to $\frac{\delta E_{NL}^{\hat v_{nl}}}{\delta n(\br)}$, let us also consider the $\frac{\delta T_s^{\hat v_{nl}}}{\delta n(\br)} - \frac{\delta T_s}{\delta n(\br)}$, a term which is more challenging in nature. This is because the NLPP and LPP KS orbitals are generally different and would not yield the same kinetic energy unless the wavefunctions coincide beyond a radial cutoff \cite{moldabekov2024nonlocal}. 
We note that PPs are derived to maintain the Coulomb tail behavior beyond a certain cutoff radius, which we shall indicate by $R_{cut}^{Coul}$.  \eqn{eq:inv} shows us that the local parts of the NLPP should be equal to the LPP beyond the larger of the $R_{cut}^{Coul}$ among NLPP and LPP, ensuring that $\Delta v_{LPP}(\br)=0$ beyond this cutoff radius. 

Unfortunately, there is generally no guarantee that $R_{cut}^{Coul}$ for the LPP KS system will be small. However, early work \cite{starkloff1977local}, pioneering work by Mi {\it et al.} \cite{mi2016first}, as well as more recent studies from the Huang group \cite{chi2024high}, demonstrate that for several transition metals, $R_{cut}^{Coul}$ can be made as small as typical cutoff radii used in standard pseudopotentials. 

Thus, we should expect to be able to construct LPPs for a significant set of transition metal elements. This is a quest we tackle head-on in the next sections.

\section{Simultaneously approximating $\Delta v_{LPP}(\br)$ and the OF-DFT OEP problem \label{methods}}

We now change gears as we consider approximate KEDFs, indicated by $\tilde T_s$, typically employed in OF-DFT simulations. All expressions derived thus far apply only to systems where an exact $T_s$ functional is used. Taking $\Delta v_{LPP}(\br)$ from \eqn{eq:inv} from the previous section, we can now reformulate it to also correct for approximate KEDFs as follows,
\begin{equation}
    \label{dvpp}
    \Delta v_{LPP}(\br) = \frac{\delta T_s^{\hat v_{nl}}}{\delta n(\br)} - \frac{\delta \tilde T_s}{\delta n(\br)} + \frac{\delta E_{NL}^{\hat v_{nl}}}{\delta n(\br)}.
\end{equation}

This definition clearly shows two goals for $\Delta v_{LPP}(\br)$: (1) it includes the effect of the nonlocal part of the KS-DFT PP; (2) it corrects for the fact that KEDFs are approximate, making sure that the density of the OF-DFT system matches the one of the KS-DFT system. Thus, the result of approximating $\Delta v_{LPP}(\br)$ is to simultaneously achieve the two goals.

To approximate it, we first impose it to be spherically symmetric around each ion, i.e., $\Delta v_{LPP}(\br) \equiv \Delta v_{LPP}(r)$, with $r=|\br-\mathbf{R}_I|$ for $\br$ in the vicinity of ion $I$, where $\mathbf{R}_I$ is the position of the $I$-th ion. Then, we impose the following boundary conditions that ensure smoothness and short rangedness, with $r_{cut}$ being a user defined cutoff radius,
\begin{equation} 
    \begin{cases}
        \Delta v_{LPP}(r) = 0 & \text{if } r = r_{cut},\\
        \frac{d\Delta v_{LPP}(r)}{dr}=0 & \text{if } r=r_{cut} \text{ and } r = 0.
    \end{cases}
    \label{eq:boundary}
\end{equation}

We choose a polynomial ansatz
\begin{equation}
    \Delta v_{LPP}(r)= a_{0} + a_{1} d(r) + a_{2} d(r)^2 + a_{3} d(r)^3 + a_{4} d(r)^4 + a_{5}d(r)^5
    \label{eq:poly}
\end{equation}
 where $d(r) = r - r_{cut}$. After applying the conditions above, we find $a_0$ and $a_1$ to vanish, and $a_{2} = 3a_{3} r_{cut} - 4a_{4} r_{cut}^2 +5 a_{5} r_{cut}^3$. The remaining three parameters ($a_3$, $a_4$ and $a_5$) are optimized minimizing the following cost function (exposing the dependence of the densities on the pseudopotentials),
\begin{equation}
    \Delta n = \frac{1}{2} \int \left| n_{OF}[v_{LPP}](\br) - n_{KS}[\hat v_{NLPP}](\br) \right| d\br,
    \label{eq:diff}
\end{equation}
where 
\begin{equation}
 n_{OF}(\br) = \arg\min_{n}\left\{ \tilde T_s [n] + E_{Hxc}[n] + \int v_{LPP}(\br)n(\br)d\br \right\}
    \label{eq:ofden}
\end{equation}
and $n_{KS}(\br)$ is the KS-DFT pseudodensity arising from a particular choice of nonlocal PP. The minimization effectively, yet approximately, implements OEP because the loss function is minimum when the OF-DFT density is close to the KS-DFT pseudodensity.

Important aspects to note are that $\Delta v_{LPP}(\br)$ is: (1) short ranged around each ion, as it is only nonzero for $0\leq |\br-\mathbf{R}_I| \leq r_{cut}$; (2) of spherical symmetry around each ion. Thus, the resulting LPP, given by $v_{loc}^{\hat v_{nl}}(r)+\Delta v_{LPP}(r)$ maintains the correct Coulomb decay in agreement with the target NLPP. Choosing an appropriate valence configuration for a pseudoatom is a complex task and has been addressed extensively in the pseudopotential literature. In this work, we simply adopt the valence configuration and the number of valence electrons from the corresponding target KS-DFT pseudopotential. As shown in Table S1, for some elements the number of valence pseudo-electrons (\texttt{zval}) differs between the two types of pseudopotentials considered. This difference arises from variations in \texttt{zval} among the corresponding KS-DFT pseudopotentials. We also note that there is freedom in choosing the cutoff radii ($r_{cut}$). We use the value of the largest of the $R_{cut}^{nl}$ of the target NLPP, see supplementary Table S1. We will see in the results section that this choice might not be the most optimal given the restrictive, polynomial ansatz for $\Delta v_{LPP}(r)$. The choice of cutoff radius will deserve further consideration in future works.

Regarding the choice of polynomial ansatz for $\Delta v_{LPP}(r)$, after testing, we found that a 5th order provides a good balance between accuracy of the LPPs and computational cost of the LPP optimizer. Reducing the order of the polynomial would result in significantly lower quality LPPs, and increasing the order would result is significant computational cost and convergence complexity for the optimizer.

Finally, we note that the proposed pseudization results in a overall smooth PP. The short-ranged polynomial is necessarily smooth because the high-order terms (beyond 5th) are not included. The Coulomb tail is recovered from the LPP of the target NLPP. Thus, the new LPPs are designed to be numerically well-behaved.

\section{Computational Details}

The KS-DFT calculations were performed using the Python package QEpy \cite{qepy}, which is interfaced with Quantum Espresso (QE) \cite{giannozzi2009quantum}. All simulations employed a Monkhorst-Pack grid \cite{PhysRevB.13.5188} of $11 \times 11 \times 11$ $k$-points and a plane wave cutoff energy of 680 eV for the wavefunctions (four times as much for the electron pseudodensity). We chose two nonlocal, ultrasoft PPs, namely, GBRV \cite{GARRITY2014446} and PSL \cite{dal2014pseudopotentials} due to their ability to retain accuracy across the periodic table while maintaining a workable plane wave cutoff. The KS-DFT calculations were run with these PPs while the OF-DFT calculations were ran with either only the local part of the PPs or with the additional $\Delta v_{LPP}(\br)$ correction discussed before.

The OF-DFT calculations were carried out using the DFTpy Python package \cite{shao2021dftpy} with a plane wave cutoff energy of 1600 eV. To compare to existing LPPs for OF-DFT simulations, we employed the High-Quality Local Pseudopotentials (HQLPP) of recent conception \cite{chi2024high}.

All calculations employed the Local Density Approximation (LDA) xc functional proposed by Perdew and Zunger \cite{perdew1981self}. 
Therefore, target results in this work are given by pseudopotential KS-DFT simulations employing the LDA xc functional. We also tested our pseudization procedure using PBE \cite{perdew1996generalized} for Ag recovering a similarly accurate pseudization as LDA (see Results section). To test the effects that different xc functionals may have on the pseudization procedure, we included Tables S10-S12 collecting EOS parameters for Ag using PBE. Testing other xc functionals will be matter of future work.

Unless otherwise stated, bulk calculations are carried out on the primitive cell of the most stable phase, the equation of state (EOS) curves were generated by varying the lattice parameter from 0.9$V_0$ to 1.2$V_0$, where $V_0$ is the equilibrium volume. From these curves, the minimum energy ($E_0$) and bulk modulus ($B_0$) were obtained by fitting the energy versus volume data to Murnaghan's equation of state \cite{PhysRevB.28.5480}. We further assess the transferability of the new pseudopotentials using a nonlocal KEDF, revHC \cite{xu2019nonlocal} which is based on the HC functional \cite{huang2010nonlocal}, for the EOS parameters.

Phonon spectra calculations were performed using experimental lattice constants and different methods for KS-DFT and OF-DFT. For KS-DFT, the phonon calculations were conducted in QE using Density Functional Perturbation Theory \cite{gonze1997first, baroni2001phonons}. For OF-DFT, the DFTpy package was used as a calculator within the Atomic Simulation Environment (ASE) Python package to compute phonons via finite differences, employing $5 \times 5 \times 5$ supercells. Due to the low computational cost of OF-DFT simulations with DFTpy \cite{shao2021efficient}, the OF-DFT phonon calculations did not add a significant overhead.

Isolated atom calculations were performed by placing one atom in a cubic cell with a side length of 15 \AA. 

\section{Results and discussion}

The new set of pseudopotentials generated by the procedure in Section \ref{methods} are termed PGBRV1.0, PPSL1.0, PGBRV0.2, and PPSL0.2 where the prepending ``P" represents the pseudization for OF-DFT, PSL and GBRV represent the reference KS NLPP used, and the digits represent the kind of KEDF used to build $\Delta v_{LPP}$. We employed the Thomas-Fermi (TF) plus von Weizs\"acker (vW) functional with 100\% vW (1.0) or 20\% vW (0.2). 

In the following sections, we analyze the performance of the new set of pseudopotentials for transition metal elements.  We first analyze the electron density of periodic systems and assess the ability of the new LPPs to reproduce the KS-DFT reference density of the most stable phase of transition metals. 
We also show that the new LPPs are applicable to several crystal phases not only the ones used for the PP development. 
We then move to analyze the electron density of isolated atoms to infer on the applicability of the PPs in combination with approximate KEDFs. 
For selected elements, we also test phonon spectra and an analysis of the electron density for cluster systems.

\subsection{Generation of the new LPPs}\label{conv_err}

The LPPs are generated according to \eqs{eq:boundary}{eq:ofden} for all transition metals. Inspired by BLPs \cite{zhou2004transferable}, we pseudized atoms in the condensed phase, choosing the most stable phases for each of the elements employing the experimental lattice constants.

The success of the pseudization is given by how well the density difference betweeen the target KS-DFT, $n_{KS}[\hat v_{NLPP}]$, and the OF-DFT, $n_{OF}[v_{LPP}]$, pseudodensities in \eqn{eq:diff} can be minimized as a function of the parameters ($a$ to $e$ in \eqn{eq:poly}) defining the LPP correction, $\Delta v_{LPP}(r)$. We choose the percentage density deviation, \(\Delta_v\), as a measure of success,
\begin{equation}
    \label{eq:percent}
    \Delta_{v} = 100\times \frac {\Delta n}{N_v}
\end{equation}
where \(N_v\) is the number of valence electrons. We note that $\Delta n \leq N_v$, thus $\Delta_{v}\leq 100$. In Supplementary Information Figure S1 we show the spatial distribution of the electron density difference between the OF and KS, using a representative cut plane and line cut for the most stable phases for Ag, Cu, and Pd. The figure validates the convergence criteria in \eqn{eq:percent}.

\begin{figure}[h!]
    \centering
    \includegraphics[width=\textwidth]{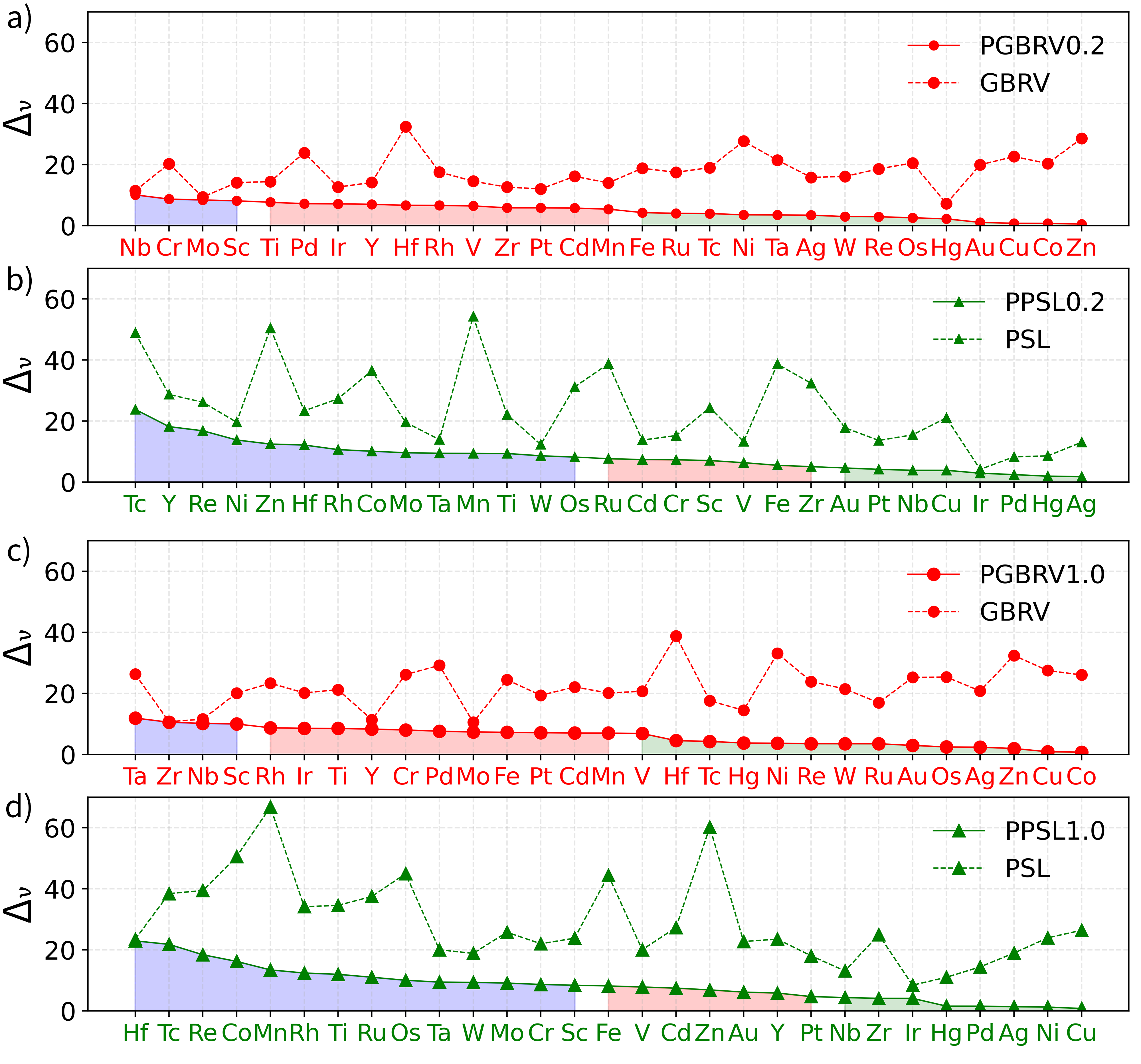}
    \caption{Percent density deviation, $\Delta_v$, for the atoms in the most stable phase of all transition metals. Green regions are elements with $\Delta_v \leq 5\%$; red  $5 < \Delta_v \leq 8\%$ and blue $\Delta_v >8\%$. The four panels are each devoted to one set of pseudopotentials (PGBRV0.2, PGBRV1.0, PPSL0.2, PPSL1.0, see text for more details).}
    \label{fig:percentage_error}
\end{figure}

Figure \ref{fig:percentage_error} provides a visual showing three sets: the color green, red, and blue; green highlights elements where the new pseudopotentials have \(\Delta_v\) lower than 5\%, red elements have \(\Delta_v\)  between 5\% and 8\%, and blue elements have \(\Delta_v\) greater than 8\%. The figure clearly shows that PGBRV LPPs generally yield pseudodensities closer to the KS reference compared to the other LPPs. We note that the pseudization dramatically improves upon the densities determined using directly the local part of the NLPP (i.e., $v_{loc}^{\hat v_{nl}}$) which show an erratic behavior with often very inadequate \(\Delta_v\) values. This is an important result as using the local part of the NLPP is a common practice in OF-DFT whenever accurate LPPs are not available\cite{fiedler2022accelerating}.

We observed that for those elements where the local part of the NLPP already provides pseudodensities $\sim15\%$ deviated from the KS-DFT reference, minimization of \eqn{eq:diff} is challenging. We observed this to be the case for Nb, Mo, and Hg for PGBRV0.2, and Zr, Nb, Y, and Mo for PGBRV1.0. 

When comparing PGBRV0.2 (PGBRV1.0) with PPSL0.2 (PPSL1.0), we observe that elements pseudized well by one are often less effectively pseudized by the other. Our analysis suggests this behavior stems from differences in the cutoff radii of the two NLPPs, which directly affect the domain of the polynomial defined in \eqn{eq:poly}. Specifically, as illustrated in supplementary Table S1 and the EOS results in Tables S2-S5, when the GBRV  \(R_{cut}^{nl}\) is larger than that of PSL (e.g., for Nb), PPSL0.2 [recalling our choice 
$r_{cut}=\max{(R_{cut}^{nl})}$ in \eqn{eq:poly}] results in smaller density deviations compared to PGBRV0.2. Conversely, for elements such as Ni, Re, Zn, Ta, and W, the smaller GBRV \(R_{cut}^{nl}\) yields more accurate pseudization with PGBRV0.2 than with PPSL0.2. Similarly, PGBRV1.0 pseudopotentials outperform PPSL1.0 for Re and W, while PPSL1.0 provides better results for Zr. These observations underscore the significance of \(r_{cut}\) in defining LPPs, suggesting that the optimal choice of cutoff radius should be element-specific and optimized.

In sum, PGBRV0.2 and PGBRV1.0 LPPs provide a better pseudization than PPSL0.2 and PPSL1.0.  Elements successfully pseudized by PGBRV0.2 and PGBRV1.0 are Co, Cu, Zn, Os, Ni, Ag, Au, Re, W, Hg, Ru, and Tc, while for Pd, Cd, V, Mn, and Pt the pseudization is moderately successful. Nb, and Sc are consistently difficult elements for both TFvW and TF0.2vW functionals and NLPPs.

\begin{figure}[h!]
    \centering
    \includegraphics[width=1.0\linewidth]{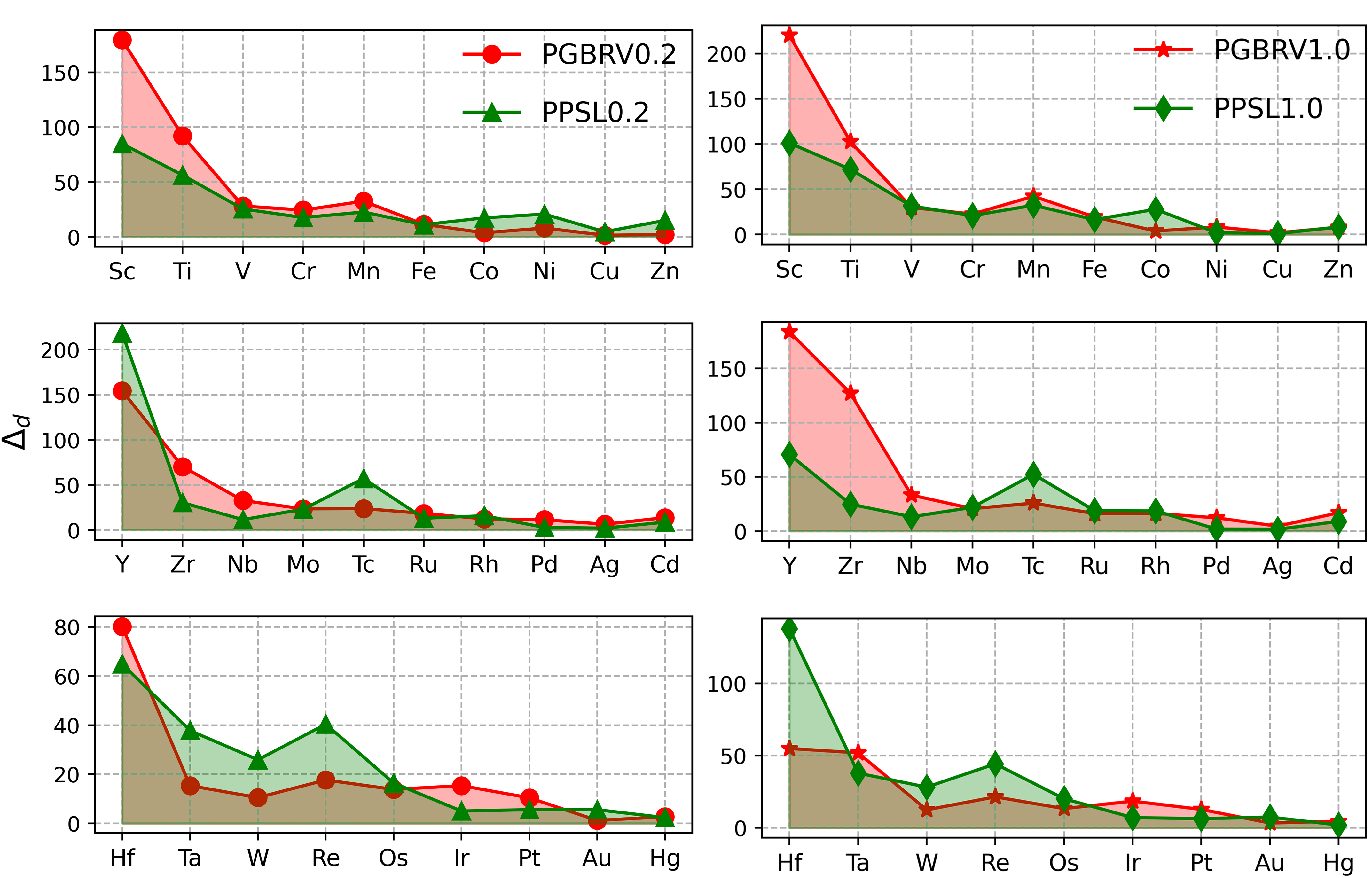}
    \caption{Density deviation per $d$ electron, $\Delta_d$. The panels compare pseudopotentials constructed using the TF0.2vW functional (left) and the TFvW functional (right) for transition metals. Red circles represent GBRV pseudopotentials, green triangles PSL.}
    \label{fig:conv_disp} 
\end{figure}

To understand (and predict) the quality of the pseudization, we consider the density displacement presented just now but we order the data by element group and split the data for the three transition metal periods. We also change slightly the plotted indicator to the density deviation relative to the number of $d$-electrons (rather than valence electrons, see Figure \ref{fig:conv_disp}), which we define as follows

\begin{equation}
    \label{eq:ddd}
    \Delta_d = 100\times \frac{\Delta n}{N_d}
\end{equation}
where $\Delta n$ is defined in \eqn{eq:diff}, and $N_d$ is the number of valence $d$-electrons. $\Delta_d$ normalizes $\Delta n$ differently from \eqn{eq:percent}. Specifically, $N_v$ includes semicore electrons while $N_d$ only includes valence $d$ electrons. Some elements may be pseudized differently by the GBRV and PSL NLPPs. For example Ag with GBRV has $4s$ and $4p$ semicore electrons (total of 8) while PSL includes no semicore electrons in the pseudodensity of this element.

A general trend arises from Figure \ref{fig:conv_disp}: $\Delta_d$ varies monotonically along the transition metal rows. That is, for late transition metals (e.g., Cu, Zn, Ag, Cd, Au, and Hg) both types of LPPs converge to similar, low $\Delta_d$ values indicating good pseudization. In contrast, early transition metals (e.g., Sc, Ti, and Y) exhibit larger displacements for both LPPs. We interpret this behavior noticing that KEDFs are known to perform better as the number of electrons increases, consistent with the exactness of the Thomas-Fermi functional in the limit of an infinite electron system \cite{constantin2011semiclassical}. 

We also observe considerable variability in results for group 3–5 transition metals, independent of the period. For example, the PGBRV pseudopotentials yield higher displacements for Sc and Ti compared to PPSL, whereas for Y, PPSL1.0 performs best, followed by PPSL0.2 and then the PGBRV LPPs. While the exact reasons for this behavior remain unclear (we note that GBRV and PSL pseudize each of Y, Sc and Ti with the same number of valence electrons), as noted before it likely relates to the different $r_{cut}$ values employed in these LPPs. For Y, GBRV uses $r_{cut}=2.2\,a_0$, while PSL uses $r_{cut}=1.8\,a_0$. This highlights the need for a follow-up study focused on optimizing $r_{cut}$ for these new pseudopotentials. Although we have adopted $r_{cut}$ values from the original NLPPs, more suitable choices may exist.

\subsection{Equation of State} \label{EOS}

From the previous analysis, we have chosen to further investigate PGBRV0.2 for a selected set of atoms such as Cu, Ag, and Pd for which PGBRV0.2 is expected to improve upon the local part of the GBRV PP (i.e., the $v_{loc}^{\hat v_{nl}}$ mentioned before). We also present Mo for which we predict the PGBRV0.2 will not provide improved electronic structure compared to the local part of the GBRV NLPP. The analysis involves computation of EOS parameters (equilibrium volume, $V_0$; equilibrium energy, $E_0$; and bulk modulus, $B_0$) for FCC, BCC, and HCP phases. We compare the PGBRV0.2 results with the local part of the GBRV NLPP and HQLPP. All calculations are carried out with TF0.2vW and the nonlocal revHC KEDFs.  The results are collected in Tables \ref{tab:Ag_eos} for Ag, \ref{tab:Cu_eos} for Cu, \ref{tab:Pd_eos} for Pd, and \ref{tab:Mo_eos} for Mo, where we show the relative error of \(V_0\) and \(B_0\) and the energy
difference compared to the most stable phase of the presented elements.
The EOS parameters from KS (LDA) are compared to experimental values in Table S9 showing that the KS-DFT benchmark results are close to experimental ground truth for lattice constants and within 30\% for bulk moduli.

Generally, we notice that PGBRV offers an improved description compared to the local part of GBRV regardless of the functional employed. In comparison to HQLPP, the PGBRV LPPs generally perform better. For example, they provide better $V_0$ and better $B_0$. This is true generally except perhaps for Mo. We notice that for Mo the $\Delta_v$ metric for pseudization did not show any improvement compared to the local part of GBRV (see panel (a) of Figure \ref{fig:percentage_error}).

Analyzing the specifics, we see that while the EOS parameters values for either TF0.2vW or revHC are not quantitative, they recover important trends for the equilibrium volume (even though it is generally underestimated by about 10-30\%) as well as the bulk modulus. 
For the latter, we note that approximate KEDFs generally struggle to produce qualitative bulk moduli. However, despite the large deviation, we can discern a generally recovered trend. For example, HCP $B_0$ is systematically larger than the FCC or BCC $B_0$. The energy ordering of the phases is also generally recovered with FCC consistently predicted to be the most stable phase. 

Even though the results presented in Tables~\ref{tab:Ag_eos}--\ref{tab:Mo_eos}, as well as in supplementary Tables S2-S5, are not quantitatively accurate, they are nonetheless reasonable. In Table S6 we show a statistical summary of Tables S2-S5, where the  Mean Absolute Relative Error (MARE) values obtained for \(V_0\) and \(B_0\) using the original GBRV and PSL pseudopotentials indicate high deviations between the OF-DFT and KS-DFT results. In contrast, the new local pseudopotentials (PGBRV0.2 and PPSL0.2) significantly reduce the MARE for $V_0$ to 28–37\%, corresponding to a significant improvement in accuracy relative to employing the local part of the KS-DFT pseudopotentials. Although the $B_0$ values remain less accurate (MARE \(\approx\)55–60\%),  all materials are predicted to be bound. This results are nontrivial for OF-DFT calculations\cite{rios2024effective,witt2021random}, and as previously noted, the OF-DFT calculations employing the newly developed LPPs successfully capture several trends observed in the KS-DFT EOS parameters. To rationalize this outcome, we emphasize that the PGBRV0.2 LPP is designed not only to partially recover effects associated with the nonlocal component of the NLPP, but also, crucially, to locally invert the KS density in regions near the ionic cores. Consequently, using the PGBRV0.2 pseudopotentials yields electron densities of substantially higher quality than would be obtained using a generic, KEDF-agnostic LPP. A clear demonstration of this point is provided by the performance comparison with HQLPP pseudopotentials, which yield lower accuracy than PGBRV0.2. A more detailed quantitative comparison between the performance of PGBRV0.2 and HQLPP pseudopotentials is presented in Table S7, where statistical metrics are reported for the FCC, BCC, and HCP phases of Ag, Cu, Pd, and Mo using the TF0.2vW functional. When switching to the revHC functional (Table S8), the PGBRV0.2 pseudopotentials retain a similar level of accuracy (MARE \(\approx\) 20–40 \%), whereas the HQLPPs show noticeably larger deviations. These results confirm that the new pseudopotentials achieve improved consistency with respect to KS-DFT and additionally that their improved performance over using the local part of the KS-DFT PPs is retained across various KEDFs.  This discrepancy does not reflect a flaw in the HQLPP pseudopotentials, as they were explicitly constructed to best represent electron densities calculated with the exact KEDF.

\begin{table}[h!]
    \centering
    \caption{Percentage errors in the equation of state parameters for \(V_0\) and \(B_0\) with respect to KS-DFT results, and energy difference against the stable phase \(E_0\) of Ag. A dash --- indicates that the phase is found to be unbound.}
    \label{tab:Ag_eos}
    \begin{tabular}{c|c|c|ccc}
    \toprule
    \hline
    Ag 
    & Functional
    & PP 
    & \begin{tabular}{c}$V_{0}$\\ (\%)\end{tabular} 
    & \begin{tabular}{c}$E_{0}$\\(eV/atom)\end{tabular} 
    & \begin{tabular}{c}$B_{0}$\\(\%)\end{tabular} \\
    \hline
\multirow{7}{*}{\rotatebox{90}{FCC}}  
    &\multirow{1}{*}{KS}          &GBRV     & 0.00   & -4023.32   & 0.00   \\ \cline{2-6}
    &\multirow{3}{*}{TF0.2vW}     &GBRV     & +43.0 & -3719.80& -44.6 \\
    &                              &PGBRV0.2 & -20.1 &-4546.42 & -22.3 \\
    &                              &HQLPP    & +34.7 &-4054.91 & -50.4 \\ \cline{2-6}
    &\multirow{3}{*}{revHC}        &GBRV     & +58.1 &-3731.58 & -56.8 \\
    &                              &PGBRV0.2 & -13.3 &-4380.85 & -12.2 \\
    &                              &HQLPP    & +22.6 &-3998.92 & -50.4 \\
\hline
\multirow{7}{*}{\rotatebox{90}{HCP}}  
    &\multirow{1}{*}{KS}          &GBRV     & 0.00    & 0.01 & 0.00   \\ \cline{2-6}
    &\multirow{3}{*}{TF0.2vW}     &GBRV     & +35.8  & 0.14 & -59.5 \\
    &                              &PGBRV0.2 & -17.2 & 0.04 & -30.1 \\
    &                              &HQLPP    & +33.9 & 0.01 & -66.5 \\ \cline{2-6}
    &\multirow{3}{*}{revHC}        &GBRV     & ---   & ---  & ---   \\
    &                              &PGBRV0.2 & -14.1 & 0.01 & -27.7 \\  
    &                              &HQLPP    & +19.0 & 0.01 & -35.8 \\
\hline			
\multirow{7}{*}{\rotatebox{90}{BCC}}  
    &\multirow{1}{*}{KS}          &GBRV     & 0.00   & 0.04 & 0.00   \\ \cline{2-6}
    &\multirow{3}{*}{TF0.2vW}     &GBRV     & +42.8 & 0.05 & -45.7 \\
    &                              &PGBRV0.2 & -20.7 &0.04 & -25.4 \\
    &                              &HQLPP    & +34.7 &0.00 & -49.3 \\ \cline{2-6}
    &\multirow{3}{*}{revHC}        &GBRV     & ---   & ---   & ---   \\
    &                              &PGBRV0.2 & -14.4 & 0.0 & -10.1 \\  
    &                              &HQLPP    & +34.5 & -0.02  & -55.8 \\
    \hline
    \bottomrule
    \end{tabular}
\end{table}

\begin{table}[h!]
    \centering
    \caption{Percentage errors in the equation of state parameters for \(V_0\) and \(B_0\) with respect to KS-DFT results, and energy difference against the stable phase \(E_0\) of Cu. A dash --- indicates that the phase is found to be unbound.}
    \label{tab:Cu_eos}
    \begin{tabular}{c|c|c|ccc}
    \toprule
    \hline
    Cu 
    & Functional
    & PP 
    & \begin{tabular}{c}$V_{0}$\\ (\%)\end{tabular} 
    & \begin{tabular}{c}$E_{0}$\\(eV/atom)\end{tabular} 
    & \begin{tabular}{c}$B_{0}$\\(\%)\end{tabular} \\
    \hline
\multirow{7}{*}{\rotatebox{90}{FCC}}  
    &\multirow{1}{*}{KS}          &GBRV     & 0.00    & -5490.58  & 0.00   \\ \cline{2-6}
    &\multirow{3}{*}{TF0.2vW}     &GBRV     & +81.0   & -4613.32  & -63.5 \\
    &                              &PGBRV0.2 & +11.5  & -6947.80  & -56.1 \\
    &                              &HQLPP    & +78.5  &  -5599.68 & -73.5 \\ \cline{2-6}
    &\multirow{3}{*}{revHC}        &GBRV     & +115.0 & -4625.80	 & -66.7 \\
    &                              &PGBRV0.2 & +27.2  & 6979.97	  & -75.7 \\
    &                              &HQLPP    & ---     & ---        & ---    \\
\hline
\multirow{7}{*}{\rotatebox{90}{HCP}}  
    &\multirow{1}{*}{KS}          &GBRV     & 0.00     &  0.02   & 0.00   \\ \cline{2-6}
    &\multirow{3}{*}{TF0.2vW}     &GBRV     & ---      & ---      & ---    \\
    &                              &PGBRV0.2 & +8.8   &  0.01	 & -55.8 \\
    &                              &HQLPP    & +103.2 & 0.02	  & -94.4 \\ \cline{2-6}
    &\multirow{3}{*}{revHC}        &GBRV     & ---     &  ---      & ---    \\
    &                              &PGBRV0.2 & -8.3   & -0.038	  & -56.7 \\  
    &                              &HQLPP    & +84.4  & -5483.58 & -52.8 \\
\hline			
\multirow{7}{*}{\rotatebox{90}{BCC}}  
    &\multirow{1}{*}{KS}          &GBRV     & 0.00     &  0.05 & 0.00   \\ \cline{2-6}
    &\multirow{3}{*}{TF0.2vW}     &GBRV     & +81.3   &  0.02 & -64.5 \\
    &                              &PGBRV0.2 & +11.1  &  0.01 & -55.9 \\
    &                              &HQLPP    & +116.3 & 0.00  & -98.4 \\ \cline{2-6}
    &\multirow{3}{*}{revHC}        &GBRV     & +144.0 &  0.07 & -86.6 \\
    &                              &PGBRV0.2 & +24.8  & 0.00	 & -73.7 \\  
    &                              &HQLPP    & ---     & ---    & ---    \\
    \hline
    \bottomrule
    \end{tabular}
\end{table}

\begin{table}[h!]
    \centering
    \caption{Percentage errors in the equation of state parameters for \(V_0\) and \(B_0\) with respect to KS-DFT results, and energy difference against the stable phase \(E_0\) of Pd. A dash --- indicates that the phase is found to be unbound.}
    \label{tab:Pd_eos}
    \begin{tabular}{c|c|c|ccc}
    \toprule
    \hline
    Pd 
    & Functional
    & PP 
    & \begin{tabular}{c}$V_{0}$\\ (\%)\end{tabular} 
    & \begin{tabular}{c}$E_{0}$\\(eV/atom)\end{tabular} 
    & \begin{tabular}{c}$B_{0}$\\(\%)\end{tabular} \\
    \hline
\multirow{7}{*}{\rotatebox{90}{FCC}}  
    &\multirow{1}{*}{KS}          &GBRV     & 0.00   &  -2704.79& 0.00   \\ \cline{2-6}
    &\multirow{3}{*}{TF0.2vW}     &GBRV      & +68.8 &  -2336.16& -69.7 \\
    &                              &PGBRV0.2 & +14.5 & -2929.42 & -61.8 \\
    &                              &HQLPP    & +55.2 &-3544.12  & -76.3 \\ \cline{2-6}
    &\multirow{3}{*}{revHC}        &GBRV     & +51.6 &-2343.41	 & -51.3 \\
    &                              &PGBRV0.2 & -25.2 &-2927.97	 & +7.9  \\  
    &                              &HQLPP    & +31.5 &-3488.77  & -71.5 \\

\hline
\multirow{7}{*}{\rotatebox{90}{HCP}}  
    &\multirow{1}{*}{KS}          &GBRV     & 0.00   &  0.03 & 0.00   \\ \cline{2-6}
    &\multirow{3}{*}{TF0.2vW}     &GBRV     & +58.8 &  0.20 & -84.6 \\
    &                              &PGBRV0.2 & +11.0 & 0.00  & -59.2 \\
    &                              &HQLPP    & +43.6 &.01	  & -68.9 \\ \cline{2-6}
    &\multirow{3}{*}{revHC}        &GBRV     & +45.2 &0.061	 & -66.3 \\
    &                              &PGBRV0.2 & -26.4 &-0.006 & -13.9 \\  
    &                              &HQLPP    & ---    & ---  & ---   \\

\hline			
\multirow{7}{*}{\rotatebox{90}{BCC}}  
    &\multirow{1}{*}{KS}          &GBRV     & 0.00    & 0.05  & 0.00   \\ \cline{2-6}
    &\multirow{3}{*}{TF0.2vW}     &GBRV     & +67.7  & 0.02  & -67.3 \\
    &                              &PGBRV0.2 & +14.1 & 0.02  & -60.5 \\
    &                              &HQLPP    & +52.4 &0.00   & -73.2 \\ \cline{2-6}
    &\multirow{3}{*}{revHC}        &GBRV     & +47.8 &-0.08  & -43.6 \\
    &                              &PGBRV0.2 & -26.6 &0.00	 & +14.1 \\  
    &                              &HQLPP    & +22.0 &0.00	 & -60.9 \\
    \hline
    \bottomrule
    \end{tabular}
\end{table}

\begin{table}[h!]
    \centering
    \caption{Percentage errors in the equation of state parameters for \(V_0\) and \(B_0\) with respect to KS-DFT results, and energy difference against the stable phase \(E_0\) of Mo. A dash --- indicates that the phase is found to be unbound.}
    \label{tab:Mo_eos}
    \begin{tabular}{c|c|c|ccc}
    \toprule
    \hline
    Mo 
    & Functional
    & PP 
    & \begin{tabular}{c}$V_{0}$\\ (\%)\end{tabular} 
    & \begin{tabular}{c}$E_{0}$\\(eV/atom)\end{tabular} 
    & \begin{tabular}{c}$B_{0}$\\(\%)\end{tabular} \\
    \hline
\multirow{7}{*}{\rotatebox{90}{BCC}}  
    &\multirow{1}{*}{KS}          &GBRV     & 0.00   &  -1897.97   & 0.00   \\ \cline{2-6}
    &\multirow{3}{*}{TF0.2vW}     &GBRV     & +22.3 &  -2140.32   & -72.4 \\
    &                              &PGBRV0.2 & +38.1 & -1990.97    & -74.8 \\
    &                              &HQLPP    & +40.9 &-1906.13	 & -77.3 \\ \cline{2-6}
    &\multirow{3}{*}{revHC}        &GBRV     & -3.2  & -2145.08    & -47.9 \\
    &                              &PGBRV0.2 & +19.7 &-1994.48	 & -62.2 \\
    &                              &HQLPP    & -11.3 &-1863.92	 & -9.8 \\
\hline
\multirow{7}{*}{\rotatebox{90}{FCC}}  
    &\multirow{1}{*}{KS}          &GBRV     & 0.00   &  0.43  & 0.00   \\ \cline{2-6}
    &\multirow{3}{*}{TF0.2vW}     &GBRV     & +20.4 &  -0.02 & -69.6 \\
    &                              &PGBRV0.2 & +34.7 & 0.04   & -74.5 \\
    &                              &HQLPP    & +38.4 &0.00	  & -74.1 \\ \cline{2-6}
    &\multirow{3}{*}{revHC}        &GBRV     & -2.3  & 0.11	  & -45.6 \\
    &                              &PGBRV0.2 & +19.9 &0.00  & -59.3 \\
    &                              &HQLPP    & -11.8 &0.01	  & -5.7 \\
\hline
\multirow{7}{*}{\rotatebox{90}{HCP}}  
    &\multirow{1}{*}{KS}          &GBRV     & 0.00   & 0.44 & 0.00   \\ \cline{2-6}
    &\multirow{3}{*}{TF0.2vW}     &GBRV     & +17.2 & 0.01 & -69.6 \\
    &                              &PGBRV0.2 & +30.0 &0.04  & -73.7 \\
    &                              &HQLPP    & +36.9 &0.00  & -78.2 \\ \cline{2-6}
    &\multirow{3}{*}{revHC}        &GBRV     & -1.5  &0.02  & -56.4 \\
    &                              &PGBRV0.2 & +18.5 &0.0  & -65.7 \\
    &                              &HQLPP    & -12.1 &0.0  & -23.4 \\
    \hline
    \bottomrule
    \end{tabular}
\end{table}

\subsection{Phonons}\label{phonon}
\begin{figure}[htp]
    \centering
    \includegraphics[width=\textwidth]{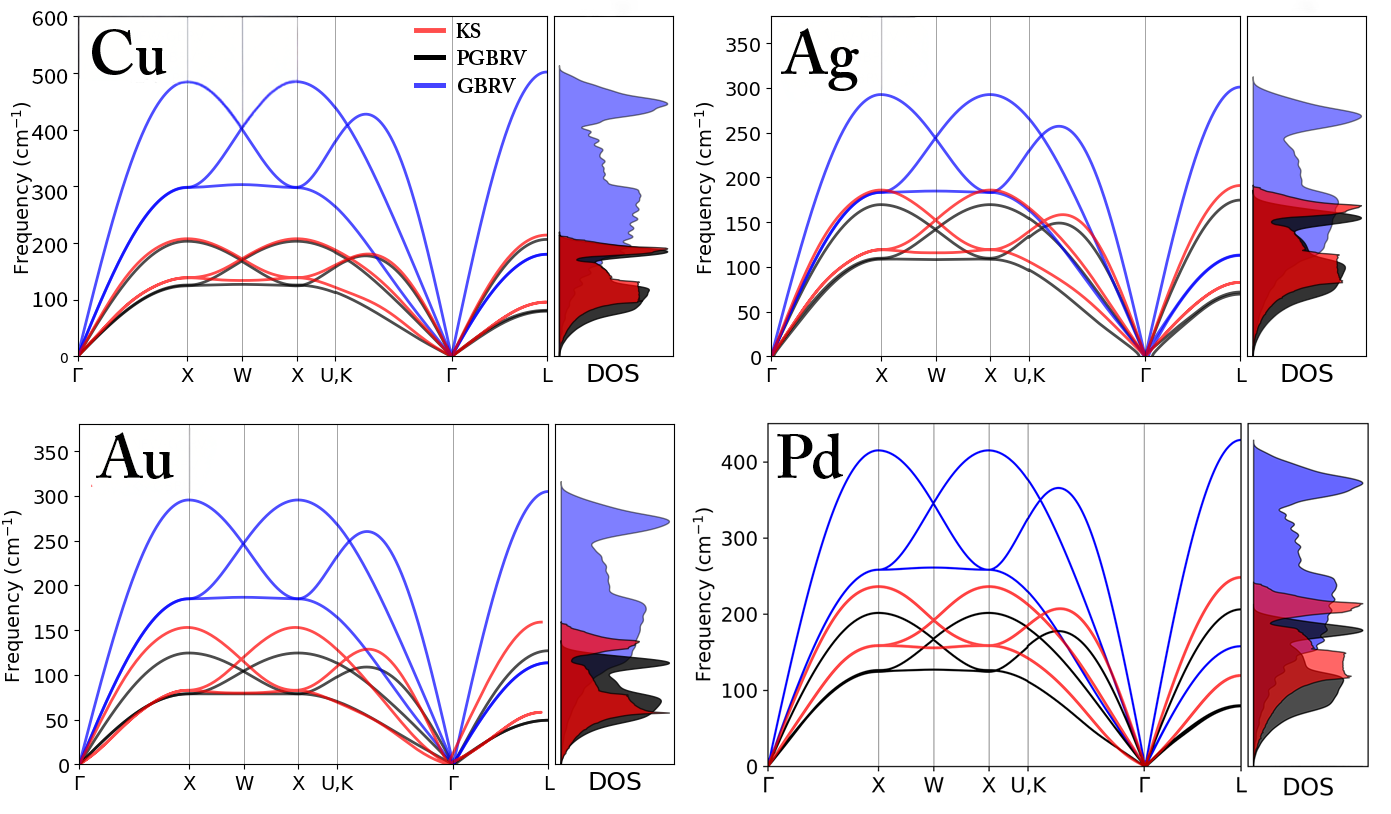}
    \caption{Phonon spectra of Cu, Ag, Au, and Pd calculated with OF-DFT using TF0.2vW with PGBRV0.2 LPPs (black lines) and the local part of the GBRV NLPPs (blue lines). Reference KS-DFT are also included (red line).}
    \label{fig:phon_Au_Cu_Ag}
\end{figure}

Figure \ref{fig:phon_Au_Cu_Ag} shows the phonon spectra for the noble metals in their FCC phase (Cu, Ag, and Au) as well as Pd also in the FCC phase calculated with TF0.2vW and two LPPs: the local part of GBRV and PGBRV0.2. The reference KS-DFT with GBRV NLPP is also included. Despite noticing a general underestimation, phonon dispersions produced by PGBRV0.2 are in good agreement with KS-DFT results and are seen to improve dramatically on the results from using the LPP from the local part of the GBRV NLPP.

From the previous sections we found that although Pd is an element that does not reach high convergence in its pseudization, we found PGBRV0.2 still delivered physical EOS parameters. For this reason, we show its phonon spectrum in the last panel of Figure \ref{fig:phon_Au_Cu_Ag}. We note that Pd phonons computed with the PGBRV0.2 LPP are once again much improved compared to using the local part of the GBRV NLPP. In the supplementary materials we show the phonon dispersion of Mo and Zr where the local part of GBRV produces several imaginary frequencies, PGBRV0.2 does not. Although for these elements the comparison with KS-DFT is not as clean as for the noble metals or Pd, the improved phonon dispersion of PGBRV0.2 vs local part of GBRV is still quite clear. These observations are consistent with our analysis of the pseudization in section \ref{conv_err}, Zr is the worst performer.
 

\subsection{Compatibility of KEDF/LPP combinations} \label{isolated}

\begin{figure}[htp]
    \centering
    \includegraphics[width=1.0\linewidth]{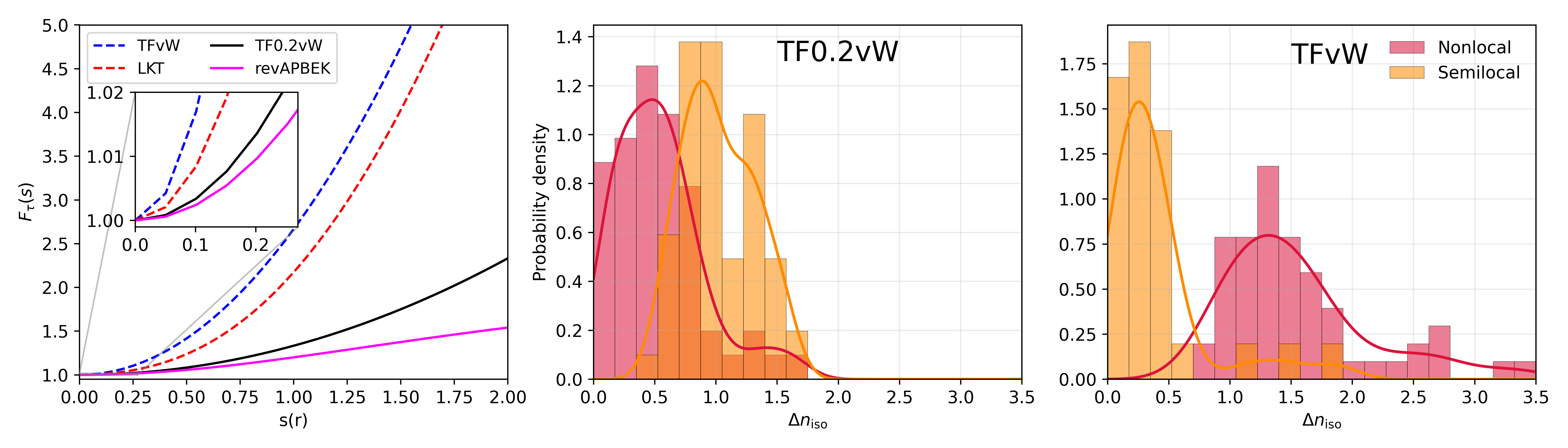}
    \caption{Left-most panel: enhancement factor, $F_\tau(s)$, of several GGA functionals. Right two panels: electron density deviation, $\Delta n_{iso}$, against TF0.2vW and TFvW functionals for several GGA KEDF (LKT, revAPBEk, TFvW) and nonlocal functionals (revHC and LMGP).}
    \label{fig:PE_isolated}
\end{figure}

To evaluate the compatibility and performance of the proposed LPPs with several KEDF approximants, we examine the electron density of isolated atoms between two classes of KEDFs: {semilocal (TFvW, TF0.2vW, LKT\cite{luo2020towards}, revAPBEK\cite{laricchia2011generalized}) and nonlocal (LMGP and revHC)} KEDFs in comparison with densities computed with either TF0.2vW or TFvW employing the new LPPs, see Figure \ref{fig:PE_isolated}. To better understand the results for semilocal KEDFs, we consider the enhancement factor, \(F(s)\), as a function of the reduced density gradient, \(s(\br)\). {Further details on the explicit forms of $F(s)$ for the various semilocal KEDFs (TFvW, TF0.2vW, LKT, and revAPBEK ) are provided in the Supplementary Information Section II.} The left most panel of Figure \ref{fig:PE_isolated} shows \(F(s)\) where we notice that for small values of \(s(\br)\), where the electron density is usually large LKT, revAPBEK , and TFvW have a noticeably higher $F(s)$ compared to  TF0.2vW. The enhancement factor for TF0.2vW, due the small vW prefactor of 0.2, displays a slower upward behavior compared to the other functionals. Similarly for large values of \(s(\br)\), which correspond to diffuse or tail regions of the density, the enhancement factor increases significantly for TFvW and LKT while it remains low for TF0.2vW and revAPBEK. The latter is due to its asymptotic behavior~\cite{laricchia2011generalized}. This analysis shows that LKT, revAPBEk and TFvW are likely more compatible with densities form TFvW in conjunction with PGBRV1.0 and PPSL1.0 than with densities from TF0.2vW in conjuction with PGBRV0.2 and PPSL0.2. 

The two histograms of $\Delta n_{\text{iso}}$ (defined in \eqn{eq:diff} bit applied to isolated atoms) in Figure \ref{fig:PE_isolated} clearly show that nonlocal functionals yield densities closer to TF0.2vW (with PGBRV0.2) with an average deviation of $0.5e$. In contrast, these functionals against TFvW (with PGBRV1.0) show an average deviation of $1.4e$. Semilocal functionals show an opposite behavior deviating by $1.0e$ from TF0.2vW and a meager $0.25e$ for TFvW.

Overall, these results emphasize the importance of internal consistency between the chosen LPP and the KEDF, particularly when applying OF-DFT to systems characterized by large density gradients or strong localization effects. This compatibility issue between KEDF and LPP was seen as a negative aspect by some \cite{zhou2004transferable}. However, in the context of this work, we see that already the two functionals considered for LPP construction, TFvW and TF0.2vW, form an excellent platform or ``basis set'' of KEDFs to be able to accommodate an array of functionals. Should one require a functional-specific LPP, the option to follow the LPP construction workflow presented here is always available and achievable using the software provided in the supplementary information document.

\subsection{Clusters}
In this section, we evaluate the performance of the new PGBRV0.2 LPPs in describing the electronic density of finite systems. Specifically, we study Pd$_{13}$, Ag$_{20}$, and Pt$_{10}$ clusters (for PPSL see supplementary Tables S6-S8). The electronic density difference, $\Delta n$, is computed as in \eqn{eq:diff}. From the results, we observe that the electron density calculated with the PGBRV0.2 LPPs is in generally good agreement with KS electron density, showing an error of 10\% or less. 

The best performers are LMGP for Pd$_{13}$ (7.3\% deviation), LKT for Ag$_{20}$ (3.8\% deviation) and TF0.2vW for Pt$_{10}$ (6.1\% deviation). When averaging over all three clusters, the TF0.2vW functional yields the smallest overall electron density difference ($\Delta n/N_v = 0.059$), thus representing the best overall performer in this set, followed closely by revHC (0.064), LMGP (0.075), LKT (0.085) and lastly TFvW (0.105). The results also show that the electron density is always improved (up to 79\% improvement!) compared to using the local part of the GBRV NLPP. 

\begin{table}[htp]
\caption{Relative density difference (against KS) for Pd$_{13}$. The OF-DFT electron density is computed with the local part of the GBRV NLPP as well as with PGBRV0.2. Improvement (\%) indicates the percent improvement of the PGBRV LPP compared to the local part of GBRV.}
\label{tab:Pd_13}
\centering
\begin{tabular}{l|c|c|c|c}
\toprule
System& KEDF & PP & \(\Delta n/N_v\) & Improvement (\%)\\ \hline
 \multirow{10}{*}{Pd$_{13}$}  & \multirow{2}{*}{TF0.2vW}   &GBRV      & 0.291 & \multirow{2}{*}{75} \\
                             &                             &PGBRV0.2  & 0.074 &  \\ \cline{2-5}
                             & \multirow{2}{*}{TFvW}       &GBRV      & 0.291 & \multirow{2}{*}{56} \\ 
                             &                             &PGBRV0.2  & 0.128 &   \\ \cline{2-5}
                             & \multirow{2}{*}{LKT}        &GBRV      & 0.268 & \multirow{2}{*}{58}  \\
                             &                             &PGBRV0.2  & 0.113 &  \\ \cline{2-5}
                             & \multirow{2}{*}{revHC}      &GBRV      & 0.240 & \multirow{2}{*}{69}\\
                             &                             &PGBRV0.2  & 0.074 & \\ \cline{2-5}
                             &  \multirow{2}{*}{LMGP}      &GBRV      & 0.221 & \multirow{2}{*}{67} \\
                             &                             & PGBRV0.2 & 0.073 &  \\ 
\bottomrule
\end{tabular}
\end{table}

\begin{table}[htp]
\caption{Relative density difference (against KS) for Ag$_{20}$. Other details as in Table \ref{tab:Pd_13}.}
\label{tab:Ag_20}
\centering
\begin{tabular}{l|c|c|c|c}
\toprule
System& KEDF & PP & \(\Delta n/N_v\) & Improvement (\%)\\ \hline
 \multirow{10}{*}{Ag$_{20}$} & \multirow{2}{*}{TF0.2vW}    &GBRV      & 0.182 & \multirow{2}{*}{77}\\
                             &                             &PGBRV0.2  & 0.042 &  \\ \cline{2-5}
                             & \multirow{2}{*}{TFvW}       &GBRV      & 0.182 & \multirow{2}{*}{67} \\ 
                             &                             &PGBRV0.2  & 0.059 & \\ \cline{2-5}
                             & \multirow{2}{*}{LKT}        &GBRV      & 0.164 & \multirow{2}{*}{77}\\
                             &                             &PGBRV0.2  & 0.038 & \\ \cline{2-5}
                             & \multirow{2}{*}{revHC}      &GBRV      & 0.128 & \multirow{2}{*}{63}\\
                             &                             &PGBRV0.2  & 0.047 & \\ \cline{2-5}
                             &  \multirow{2}{*}{LMGP}      &GBRV      & 0.107 & \multirow{2}{*}{22}\\
                             &                             & PGBRV0.2 & 0.084 &  \\ 
\bottomrule
\end{tabular}
\end{table}

\begin{table}[htp]
\caption{Relative density difference (against KS) for Pt$_{10}$. Other details as in Table \ref{tab:Pd_13}.}
\label{tab:Pt_10}
\centering
\begin{tabular}{l|c|c|c|c}
\toprule
System& KEDF & PP & \(\Delta n/N\) & Improvement (\%)\\ \hline
 \multirow{10}{*}{Pt$_{10}$} & \multirow{2}{*}{TF0.2vW}    &GBRV      & 0.290& \multirow{2}{*}{79} \\
                             &                             &PGBRV0.2  & 0.061& \\ \cline{2-5}
                             & \multirow{2}{*}{TFvW}       &GBRV      & 0.290& \multirow{2}{*}{56}  \\ 
                             &                             &PGBRV0.2  & 0.128&  \\ \cline{2-5}
                             & \multirow{2}{*}{LKT}        &GBRV      & 0.277& \multirow{2}{*}{62} \\
                             &                             &PGBRV0.2  & 0.105& \\ \cline{2-5}
                             & \multirow{2}{*}{revHC}      &GBRV      & 0.247& \multirow{2}{*}{71} \\
                             &                             &PGBRV0.2  & 0.071& \\ \cline{2-5}
                             &  \multirow{2}{*}{LMGP}      &GBRV      & 0.241& \multirow{2}{*}{72}  \\
                             &                             & PGBRV0.2 & 0.067& \\ 
\bottomrule
\end{tabular}
\end{table}

\section{Conclusions}
In this study, we developed new sets of pseudopotentials for OF-DFT calculations targeting transition metal elements. Transition metals have historically been excluded from OF-DFT applications due to the challenges of constructing suitable local pseudopotentials. The core idea of this work is to adapt existing KS-DFT pseudopotentials (specifically, PSL and GBRV) by imposing that OF-DFT calculations with approximate noninteracting kinetic energy functionals reproduce, as closely as possible, the electron density from KS-DFT calculations of the element’s most stable condensed phase. This approach effectively performs a KS-to-OF inversion near the ion cores. {Our work opens the possibility to investigate large-scale systems and nanostructures composed of transition metals at low cost (linear scaling with low prefactor). However, we also warn that simulations based on OF-DFT for transition metals should be considered highly experimental and not yet ready for broad adoption.} 

\begin{figure}[htp]
    \centering
    \includegraphics[width=\textwidth]{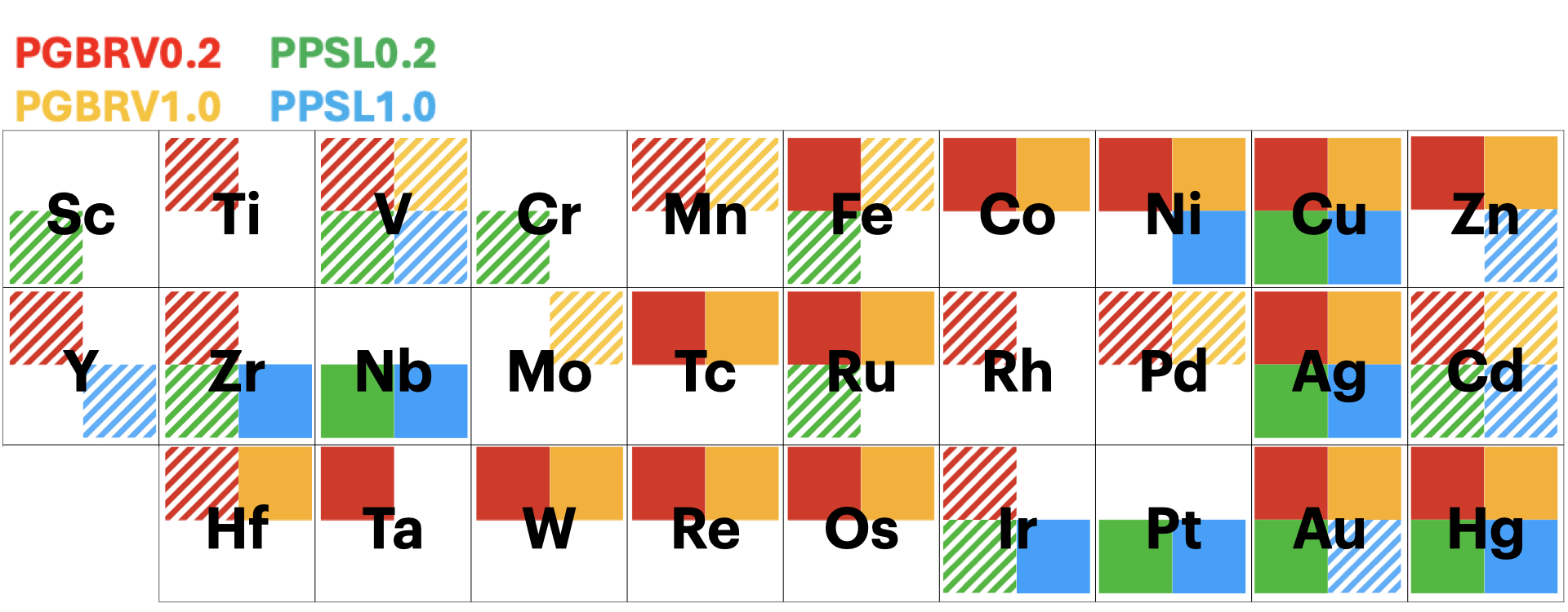}
    \caption{Recommendations for using the proposed pseudopotentials. The color map illustrates the outcome of the pseudization procedure. Solid colors represent elements for which the pseudopotential yields accurate results; hatched areas indicate lower-quality pseudization; and uncolored elements correspond to pseudopotentials we consider inaccurate. These recommendations are based on three sets of data: the electron density in the bulk system, the electron density of the isolated atom, and equation of state parameters.}
    \label{fig:periodic_table}
\end{figure}

Our convergence analysis shows that GBRV-based pseudopotentials generally outperform PSL pseudopotentials for most transition metals. Figure~\ref{fig:periodic_table} summarizes our recommendations for each element: solid colors indicate successful pseudization, while hatched areas denote intermediate accuracy (see Figure caption for details). 

We conclude with a note of caution: there are still major challenges for applying OF-DFT methods to transition metals. While pseudopotentials are increasingly available (see, e.g., Refs.\citenum{mi2016first,chi2024high,xu2022nonlocal} and this work), the overall accuracy of existing physics-based OF-DFT methods is still insufficient for the widespread use of OF-DFT with these important elements.

\section{Acknowledgements}
We thank ICTP and ASESMA (\url{https://www.asesma.org}) for promoting collaboration between the Pavanello group and EO, WFIE, SJO and AO. This work was supported by grants from the U.S. National Science Foundation under grant agreements CHE-2154760, OAC-2321103. The authors acknowledge the Office of Advanced Research Computing (OARC) at Rutgers, The State University of New Jersey for providing access to the Amarel cluster and associated research computing resources that have contributed to the results reported here. URL: \url{https://it.rutgers.edu/oarc}

\bibliography{ref}

\end{document}